\documentclass[reprint,aps,prb,floatfix,showpacs]{revtex4-1}
\usepackage{epsf}
\usepackage{epsfig}
\usepackage{float}
\usepackage{graphicx}
\usepackage{amssymb}
\usepackage{amsmath}
\usepackage{comment}
\usepackage{color}
\usepackage{setspace}
\usepackage{comment}

\def\ep{{\epsilon}}

\def\k{{{\bf k}}}
\def\om{{\omega}}
\def\CPA{\scriptscriptstyle CPA}

\def\non{{\nonumber}}
\def\g{{\bf{g}}}

\def\be{\begin{equation}}
\def\ee{\end{equation}}
\def\ber{\begin{eqnarray}}
\def\eer{\end{eqnarray}}

\def\eps{{\epsilon}}
\def\g0{{\gamma_0}}

\def\Im{{\mbox{Im}}}

\def\prb{Phys.\ Rev.\ B }
\def\prl{Phys.\ Rev.\ Lett.\ }

\def\JPCM{J.\ Phys.\ Condens.\ Matter }

\def\etal{{\it et.\ al} }

\def\RMP{Rev.\ Mod.\ Phys.\ }
\def\JPSJ{J.\ Phys.\ Soc.\ Japan}

\newcommand\bear{\begin{eqnarray}}
\newcommand\eear{\end{eqnarray}}
\newcommand\bea{\begin{align}}
\newcommand\ena{\end{align}}
\begin{document}

\title{Kondo-hole substitution in heavy fermions: dynamics and transport}
\author{Pramod Kumar}
%\affiliation{Theoretical Sciences Unit, Jawaharlal Nehru Centre for Advanced Scientific Research, \\
%Jakkur, Bangalore 560 064, India.}
\author{N.\ S.\ Vidhyadhiraja}
\altaffiliation[Also at]{-Department of Physics and Astronomy, Louisiana State University, Baton Rouge 70808, USA}
\email{raja@jncasr.ac.in}
\homepage{http://www.jncasr.ac.in/raja}
\affiliation{Theoretical Sciences Unit, Jawaharlal Nehru Centre for Advanced Scientific Research, \\
Jakkur, Bangalore 560 064, India.}

\begin{abstract}
          Kondo-hole substitution is a unique probe for exploring
the interplay of interactions, f-electron dilution and disorder in heavy fermion materials. Within the diluted periodic Anderson model, we investigate the changes in single-particle dynamics as well as response functions, as a function of
Kondo hole concentration ($x$) and temperature.  
We show that the spectral weight transfers due to Kondo hole substitution has characteristics that are different from those induced by temperature; The dc resistivity crosses over from a highly non-monotonic form with a coherence peak in the $x\rightarrow 0$ limit to a monotonic single-impurity like form that saturates at low temperature.  The thermopower exhibits a characteristic maximum as a function of temperature, the value of which changes sign with increasing $x$, and its location is shown to correspond to a low
energy scale of the system. The Hall coefficient also changes sign with increasing $x$ at zero temperature and is highly temperature dependent for all $x$. As $x$ is increased beyond a certain $x_c$, the Drude peak and the mid-infrared peak in the optical conductivity vanish almost completely; A peak in the optical scattering rate melts and disappears eventually. We discuss the above-mentioned changes in the properties
in terms of a crossover from coherent, Kondo lattice behaviour to single
impurity like, incoherent behaviour with increasing $x$.
A comparison  of theory with experiments carried out for the dc resistivity and the thermopower of Ce$_{1-x}$La$_x$B$_6$ yields good agreement. 
\end{abstract}


\maketitle

\section{Introduction}
\label{sec:intro_5}
Interest in the physics of heavy fermion materials~\cite{grew91,wach} has sustained for the past four decades because they display a rich variety of phenomena like lattice Kondo effect~\cite{hews}, large electron masses, quantum criticality~\cite{colm,gege}, valence fluctuation driven Kondo collapse and unconventional superconductivity~\cite{yoshi}. These phenomena arise mainly due to the presence of an active $f$-orbital which forms a very narrow band, and thus leads to strong correlations~\cite{hews,aepp,varm}. The concentration of $f$- electrons in heavy fermion alloys can be tuned by substituting non magnetic homologues; for example Lanthanum (La) can be substituted for Cerium(Ce). Various examples of such alloys are Ce$_{1-x}$La$_x$Cu$_6$~\cite{onuki}, Ce$_{1-x}$La$_x$B$_6$~\cite{sato}, Ce$_{1-x}$La$_x$Cu$_2$Si$_2$~\cite{ocko}, and Yb$_{1-x}$Lu$_x$Rh$_2$Si$_2$~\cite{kohl} etc. Substitution with non-magnetic homologue, defined as Kondo-hole (KH) type substitution, leads to a crossover from coherent lattice to incoherent single impurity behaviour. Such a crossover is reflected in dynamics and transport properties. The other kind of substitution in heavy fermions is ligand field substitution, as in CeCu$_{6-x}$Au$_x$~\cite{loh} and UCu$_{5-x}$Pd$_x$~\cite{andr}. This kind of doping in the former leads to a quantum critical point, that in turn manifests in a wide parameter space at finite temperatures and leads to anomalous properties.

 Experimentally, the changes in physical properties due to Kondo-hole type substitution are quite well known. With increasing disorder, the coherence peak in resistivity vanishes while the high temperature single-impurity Hammann form is preserved~\cite{onuki,sato}. The magnitude of the characteristic peak in the thermopower decreases with increasing disorder strength~\cite{sato}. In the extreme dilution limit, the peak even changes sign~\cite{sato}. However, these features and their detailed doping dependence is quite material-specific~\cite{sato}. The Hall coefficient, $R_H$, which is constant ($-1/ne$) with temperature for normal metals, is highly temperature and concentration dependent for heavy fermion metals. With increasing concentration of Kondo holes, the magnitude of Hall coefficient extrapolated to zero temperature ($R_H(T\rightarrow 0)$) changes sign~\cite{onuki2}.

 %Within the dCPA framework, Grenzebach {\it et al}~\cite{grenz} have used NRG, while Mutao ~\cite{mutou} has used IPT to study the resistivity and thermopower behaviour for Kondo hole and ligand field type disorder over the entire range of  $x\in [0,1]$.    

%   The aforementioned crossover from coherent to incoherent behaviour was also thoroughly investigated~\cite{grenz,kaul,fulde}. 
%In a recent work using slave-boson mean-field theory, $f$-electron dilution has been studied numerically on a square lattice with upto $20\times 20$ sites for different $n_c$ and $n_f$~\cite{kaul}. It was found the crossover occurs at an $x$ value that is dependent on the conduction electron filling, $n_c$, as $1-n_c$~\cite{kaul,fulde}. 
%The disadvantage with CPA is that inter-site coherence and coherent back scattering effects are ignored and hence Anderson localization cannot be incorporated. Statistical DMFT and typical mean field theory (TMT) are the extensions of CPA to account for above effects~\cite{mira1}.
         In the present paper, our main aim is to explore the dynamics
and transport quantities across the Kondo-hole doping induced crossover from coherent HFs to incoherent single impurity behaviour.  Theoretical work on heavy fermions (HFs) with Kondo hole substitution, modeled
 by the periodic Anderson model (PAM), has been extensive. A standard approach is to embed the coherent potential approximation (CPA)~\cite{jani,dobro,ulmke} within the dynamical mean field theory (DMFT) framework which yields a dynamical CPA (dCPA) ~\cite{grenz,mutou}. The dCPA has been employed in combination with impurity solvers such as slave boson (SB) mean field theory~\cite{mira,kaul}, numerical renormalization group (NRG)~\cite{grenz} and iterative perturbation theory (IPT)~\cite{mutou} to investigate the diluted PAM. We have derived dCPA equations using a Feenberg renormalized perturbation series and have employed the local moment approach~\cite{loga98} as an impurity solver. The comprehensive NRG work by Grenzebach et al~\cite{grenz} focused on resistivity and thermopower in the Kondo lattice limit. While our results do concur with Ref.~\cite{grenz}, in addition, we demonstrate the existence of a universal low energy scale at finite $x$ and the dependence of the crossover on $n_c$ in the resistivity.  Substitutional effects on optical conductivity, optical scattering rate and Hall coefficient have been studied in detail for the first time.
 
   We conclude that quantitative agreement with experimental results necessitates the introduction of {\em substitution dependence} into the model parameters. Experimentally measured residual resistivity per unit concentration of magnetic impurities increases with increasing Kondo hole concentration~\cite{onuki,sato}. However, in previously reported theoretical work~\cite{grenz,mutou,wemb,cost}, the residual resistivity peaks at a certain concentration value, and is not monotonic. We have found that including concentration dependence into the orbital energy of itinerant electrons correctly reproduces the known experimental trend in residual resistivity.  

      The paper is structured as follows; We first discuss the standard model for heavy fermions, i.e. the periodic Anderson model, followed by the formalism of CPA+DMFT which is needed to incorporate disorder due to Kondo hole substitution. In section~\ref{subsec:dos_5}, we present results for spectral functions, low energy scale and hybridization. In sections~\ref{subsec:res_5} and~\ref{subsec:therm_5}, we discuss resistivity and thermoelectric behaviour. In sections ~\ref{subsec:hall}, we have discussed the effects of disorder on the Hall coefficient and Hall angle. In section~\ref{subsec:opt_5}, we shift to dynamical transport quantities, namely optical conductivity and optical scattering rate.  In the final section~\ref{sec:exp_5}, we have done a detailed comparison of theoretical results with the experimental data for resistivity and thermopower in Ce$_{1-x}$La$_x$B$_6$.

\section{Model and formalism}
\label{sec:model_5}
\subsection{Periodic Anderson model}
The periodic Anderson model (PAM) is the simplest theoretical model to understand the physics of heavy fermions in various regimes. In second quantized notation, the PAM is expressed as
\be
H_{PAM}=-\sum_{\langle ij\rangle\sigma} t_{ij} \left( c^\dagger_{i\sigma} c^{\phantom{\dagger}}_{j\sigma} + {\rm h.c}\right)  + \sum_{i} H_{ii}  
\label{eq5_1}
\ee
where the local part of the Hamiltonian is $H_{ii}=\eps_{c} \sum_\sigma c^\dagger_{i\sigma} c^{\phantom{\dagger}}_{i\sigma} + \eps_{f} \sum_\sigma f^\dagger_{i\sigma} f^{\phantom{\dagger}}_{i\sigma}+
V(\sum_\sigma f^\dagger_{i\sigma} c^{\phantom{\dagger}}_{i\sigma} + {\rm h.c}) + Un_{fi\uparrow} n_{fi\downarrow}$. In this Hamiltonian (equation~\ref{eq5_1}), the first term represents kinetic energy of conduction electrons in terms of a hopping amplitude, $t$, ($\propto \frac{t^*}{\sqrt{Z_c}}$ in the limit of large co-ordination number $Z_c$). We consider the hyper-cubic lattice for which ($D_0(\epsilon)=\exp(-(\epsilon/t^*)^2)/\sqrt{\pi} t^*$) is the bare c$-$ electron density of states. The second term is diagonal in real space, and represents in sequence, the site energy for conduction electron, localised $f$-electrons, hybridization of localised and conduction electrons and the on-site Coulomb repulsion between two localised opposite spin electrons respectively.

 In order to handle Kondo hole substitution  and the consequent disorder within DMFT~\cite{geor}, we use the coherent potential approximation (CPA) which becomes exact in the limit of infinite dimensions~\cite{jani}. We outline our method  for incorporating disorder below.
\subsection{Coherent potential approximation and dynamical mean field theory}
\label{subsec:formalism_5}
We have employed Feenberg renormalized perturbation series (FRPS)~\cite{econ} for binary distribution of disorder, i.e. $P(\epsilon_i)=(1-x)\delta(\epsilon_i - \epsilon_\alpha)+x\delta(\epsilon_i - \epsilon_\beta)$ (where $\eps_i$ can be any model parameter) and derived averaged conduction and impurity Green's function for PAM. The tight binding Hamiltonian which is expressed in second quantized notation as
\be
\hat{H}=-\sum_{ij\sigma}t_{ij}c_{i\sigma}^{\dagger}c_{j\sigma}{\phantom{\dagger}}+\sum_{i\sigma}\eps_c c_{i\sigma}^{\dagger}c_{i\sigma}{\phantom{\dagger}}\label{eq5_2}
\ee
represents kinetic energy and orbital energy of a non-interacting system. The retarded non-interacting Green's function in matrix representation is given by the following equation,
\be
\textbf{g}=[\textbf{z+t}]^{-1}
\label{eq5_3}
\ee
where $z_{ij}=\delta_{ij}(\om^+ -\eps_c)$. The local Green's function for the Hamiltonian using FRPS can be written as~\cite{econ}
\be
g_{ii}=\frac{1}{\om -\eps_c -S_i\left[\{g_{jj}(\om)\}\right]}
\label{eq5_4}
\ee
where $S_i$ is a Feenberg self-energy and is a functional of local Green's functions. Specifically, it is given by the sum of all self-avoiding graphs on the lattice, where the vertices are the local (site-excluded) Green's functions, while the lines are the hopping amplitudes connecting neighbouring sites~\cite{econ}. In the limit of infinite-dimensions, the restriction of site-exclusion may be relaxed. Thus, for example, for the Bethe lattice, where the only self-avoiding closed paths would be a single-hop to a nearest neighbour, the $S(\om)$ would be a functional only of the nearest neighbour local (diagonal) Green's functions. Since $\Sigma_{\sigma}(\om)$ is diagonal in the local approximation~\cite{geor}, the Green's function in matrix representation  is given as
\be
\bf{G_{\sigma}(\om))}=[\bf{\tilde{Z}}+\bf{t}]^{-1}
\label{eq5_6}
\ee
with $\mathbf{\tilde{Z}=z-\Sigma_{\sigma}}$. The structure of equations~\ref{eq5_3} and~\ref{eq5_6} is identical and thus the Green's function with diagonal self-energy can be written as
\be
G_{ii;\sigma}(\om)=\frac{1}{\om -\eps_c -\Sigma_{i;\sigma}(\om)-S_i\left[\{G_{jj;\sigma}(\om)\}\right]}
\label{eq5_7}
\ee
where $S_i$ is exactly the same functional of local interacting Green's functions as in the non-interacting case. 
So far, we have not invoked any disorder. For a binary alloy, $P(\epsilon)=(1-x)\delta(\epsilon - \epsilon_\alpha)+x\delta(\epsilon - \epsilon_\beta)$, where every site is surrounded by a fraction $x$ of `$\alpha$' type sites and $1-x$ of `$\beta$' type. Thus in the Feenberg self-energy, since each vertex has a sum over the sites, the argument of the functional becomes a self-averaged quantity.
\be
S_{\sigma}=  S\left[G^{\CPA}_{\sigma}\right] 
\ee
and $G^{\CPA}_{\sigma}$ is the disordered averaged CPA Green's function and given as
\ber
G^{\CPA}_{\sigma}(\om)=(1-x)G^{\alpha}_{\sigma}(\om)+xG^{\beta}_{\sigma}(\om)
\label{eq5_8}
\eer
As discussed in the introduction(section~\ref{sec:intro_5}), our focus is on substitutional disorder in $f$- sites, and hence we choose $\eps_{\alpha}=\eps_{f;\alpha}$ and $\eps_\beta=\eps_{f;\beta}$. The local conduction electron ($c$-) Green's functions for `$\alpha$' type sites are given by
\be
G_{\sigma}^\alpha(\om)=\frac{1}{\om -\eps_c -\Sigma_{\sigma}^\alpha(\om)-S[G^{\CPA}_{\sigma}(\om)]}
\label{eq5_9}
\ee
and likewise for `$\beta$' type.
Here the $\Sigma_{\sigma}^{\alpha/\beta}=\frac{V^2}{\om^+ -\eps^{\alpha/\beta}_{f}-\Sigma^{f;\alpha/\beta}(\om)}$. Using equation~\ref{eq5_9} in equation~\ref{eq5_8} gives the CPA Green's function is obtained as
\begin{align}
G^{\CPA}_{\sigma}=&\frac{1-x}{\om-\eps_c -S[G^{\CPA}_{\sigma}] -\Sigma_{\sigma}^\alpha(\om)} \non \\
&+\frac{x}{\om-\eps_c -S[G^{\CPA}_{\sigma}] -\Sigma_{\sigma}^\beta(\om)}\label{eq5_11}
\end{align}
Within the LMA~\cite{vidh05,pram}, we have a two self-energy description corresponding to the two degenerate mean-field broken symmetry  solutions with self-energy $\Sigma^{A}$ and $\Sigma^{B}$ and hence the corresponding Green's functions will be $G^{\CPA;A}_{\sigma}(\om)$ and  $G^{\CPA;B}_{\sigma}(\om)$. In the paramagnetic regime every site is surrounded by an equal number of `A' and `B' type Green's functions, hence
\be
G^{\CPA}(\om)=\frac{1}{2}[G^{\CPA;A}_{\sigma}(\om) +G^{\CPA;B}_{\sigma}(\om)]
\ee
With the up/down spin symmetries of the Green's function i.e $G^A_{\sigma}=G^B_{-\sigma}$, the above equation can be written as
\be
G^{\CPA}(\om)=\frac{1}{2}[G^{\CPA}_{\sigma}(\om) +G^{\CPA}_{-\sigma}(\om)].\label{eq5_12}
\ee
We note that $G^{\CPA}(\om)$ is independent of spin and thus the Feenberg self-energy which is the functional of the nearest neighbour CPA Green's function will be independent of spin $S_{\sigma}(\om)=S(\om)$. Combining equations~\ref{eq5_11} and~\ref{eq5_12} and the condition of Kondo hole type of disorder i.e $\Sigma_{\sigma}(\om)^{\beta}=\Sigma_{-\sigma}(\om)^{\beta}=0$ (since $\eps_{f;\beta}\rightarrow \infty$ for Kondo holes), the averaged $c$- CPA Green's function can be written as
\begin{align}
G^{\CPA}_c(\om)=&\frac{(1-x)}{2}\Big[\frac{1}{\om -\eps_c -S(\om) -\Sigma_{\sigma}(\om)}\non \\
&+\frac{1}{\om -\eps_c- S(\om)-\Sigma_{-\sigma}(\om)}\Big] \non \\
&+\left[\frac{x}{\om -\eps_c -S(\om) }\right]\,.\label{eq5_13}
\end{align}
The above equations are equivalent to the CPA+DMFT equations derived previously~\cite{grenz,mutou}. Since the CPA Green's function corresponds to that of a translationally invariant system, the $c$- CPA Green's functions can also be calculated with the following Hilbert transform
\be
G^{\CPA}_c(\om)= H\left[\gamma\right] = \int^\infty_{-\infty} \frac{\rho_0(\eps)}{
\gamma(\om)-\eps} = \frac{1}{\gamma(\om)-S(\om)}
\label{eq5_14}
\ee
where $\gamma(\om)=\omega^+ -\epsilon_c-\Sigma^{\CPA}_c$. Equations~\ref{eq5_13} and~\ref{eq5_14} form a self-consistent set of equations  for $S(\om)$ if the self-energies $\Sigma_\sigma$ are known. Since ``$\beta$" type  for Kondo hole substitution does not have ``f" electron, so $G^{\CPA}_f(\om)$ will have contribution from ``$\alpha$" only and is given by 
\be
G^{\CPA}_f(\om)=(1-x)\left[\om^{+}-\ep_f-\Sigma_f(\om)-\frac{V^2}{\om^{+}-\ep_c-S(\om)}\right]^{-1}
\ee 
Finally, the local Green's functions for the $\alpha$ type $f$- and $c$- electrons are given as
\ber
G^f_{\sigma}(\omega)=\left[\om^+ - \eps_f - \Sigma^f_{\sigma}(\om) - \frac{V^2}{\om^+ -
\eps_c - S(\om)}\right]^{-1} 
\label{eq:locgf}\\
G^c_{\sigma}(\omega)=\left[\om^+ - \eps_c - S(\om) - \frac{V^2}{\om^+ -
\eps_f - \Sigma^f_{\sigma}(\om)} \right]^{-1}
\label{eq:locgc}
\eer
 Evaluation of local self-energy $\Sigma^f_{\uparrow/\downarrow}$ and the self-consistency of DMFT is carried out in the manner discussed in detail in previous works ~\cite{vidh05,pram} for the clean case. For disordered systems, the CPA Green's functions are used to evaluate transport properties which have been discussed in the next section.
\subsection{Transport formalism}
Since within DMFT, vertex corrections are absent~\cite{geor}, the single-particle Green's functions are sufficient within the Kubo formalism to obtain transport quantities such as DC resistivity and optical conductivity. The expressions have been derived previously~\cite{vidh05} for non-disordered case for a hypercubic lattice. With the inclusion of disorder at CPA level, the expressions retain the same form, but the c$-$ Green's function is replaced by the CPA Green's function. Thus the expression for the real part of optical conductivity is 
\begin{align}
\sigma(\om; T)&=\frac{\sigma_0}{2\pi^2}\int^\infty_{-\infty} \rho_0(\eps) \int^\infty_{-\infty} \,d\om^\prime \,
\frac{n_F(\om^\prime) - n_F(\om + \om^\prime)}{\om} \non \\
&D^{\CPA}_c(\eps,\om^\prime)D^{\CPA}_c(\eps,\om+\om^\prime)
\label{eq5_15}
\end{align}
where $\sigma_0=4\pi e^2 t^2 a^2 n/\hbar$ for a lattice constant $a$,
electronic charge $e$, and electron density $n$ and $D^{\CPA}_c(\eps,\om)=-\Im{G^{\CPA}_c(\eps,\om)}/\pi$. By carrying out a Kramers-Kronig transform $\sigma^\prime(\om;T)={\cal P}\int_{-\infty}^{\infty}d\om^\prime\frac{\sigma(\om^\prime)}{\om-\om^\prime}$ of the  $\sigma(\om ;T)$  we can get $\sigma^\prime(\om;T)$, and then the complex optical conductivity, ${\bar{\sigma}(\om;T)}$, can be obtained as $\sigma(\om;T)+i\sigma^\prime(\om;T)$. The optical scattering rate is defined as~\cite{basov} $M^{-1}(\om;T)={\rm Re}(1/\bar{\sigma}(\om; T))$. 

 The DC conductivity, thermopower and Hall coefficient can be  expressed in terms of Lorenz numbers as~\cite{jaime}
\[
\sigma_{\scriptstyle{DC}}=L^{11}; S=-\frac{1}{e T}\frac{L^{12}}{L^{11}};R_H=\frac{L^{21}}{L^{11}*L^{11}}
\]
The explicit expressions for $L_{ij}$'s are as follows~\cite{jaime}.
\begin{equation}
L^{11}=\frac{\sigma_0}{2\pi^2}\int^\infty_{-\infty} \rho_0(\eps)\int^\infty_{-\infty} \,d\om \,\left(-\frac{\partial n_F}{\partial \om}\right) \,{D^{\CPA}_c(\eps,\om)}^2 \label{eq5_16}
\end{equation}
\begin{equation}
L^{12}=\frac{\sigma_0}{2\pi^2}\int^\infty_{-\infty} \rho_0(\eps)\int^\infty_{-\infty} \,d\om \,\om \,\left(-\frac{\partial n_F}{\partial \om}\right) \, {D^{\CPA}_c(\eps,\om)}^2 \label{eq5_17}
\end{equation}
\begin{equation}
L^{21}=R_0\frac{\sigma_0}{2\pi^2}\int^\infty_{-\infty} \, \eps \, \rho_0(\eps) \,\int^\infty_{-\infty} \,d\om \,\left(-\frac{\partial n_F}{\partial \om}\right) \, {D^{\CPA}_c(\eps,\om)}^3 \label{eq5_18}
\end{equation}
where $R_0=\frac{2}{3} \pi e^2$.
\section{Results and discussion}
\label{result_5}
As mentioned in the introduction, our main objective in this work is to elucidate the emergence of incoherence in heavy fermions through the introduction of Kondo holes. The focal theme throughout this section will be the crossover from coherent lattice behaviour to incoherent single-impurity behaviour as a function of the concentration of Kondo holes. The manifestation of this crossover will be examined in single-particle quantities such as spectral functions, and two-particle quantities such as DC conductivity, optical conductivity, optical scattering rate, thermopower and Hall coefficient. It is well known~\cite{vidh05,grenz1}  that heavy fermions systems display such a crossover even in the clean limit with an increase in temperature from $T=0$ to beyond the lattice coherence temperature. We will examine the interplay of disorder and temperature in inducing the incoherence. The conduction band centre is fixed at $\epsilon_c=0.5$. We begin with single-particle dynamics, i.e, with the density of states and low energy scale. Next, we explore two particle static quantities and finally, we will discuss two-particle dynamical quantities.
\subsection{Density of states, and low energy scale}
\label{subsec:dos_5}
\begin{figure}[h]
\centering{
\includegraphics[scale=0.50,clip=]{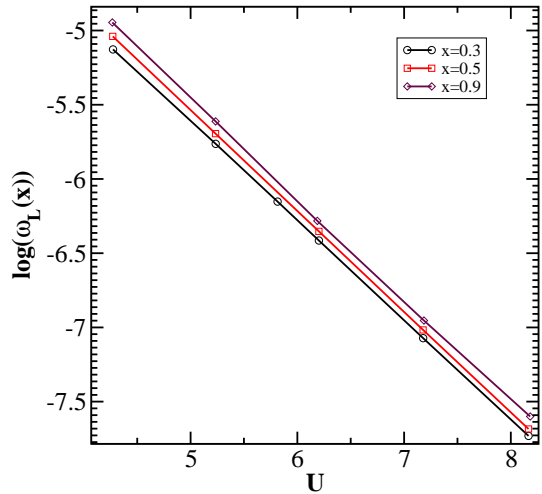}
}
\caption{(color online) Low energy scale $\om_L(x)$ varying with Coulomb intraction $U$ for different substitution values of $x$. The hybridization is chosen to be $V^2=0.4$, and the conduction band centre is at $\epsilon_c=0.5$.
}
\label{fig:unv1}
\end{figure}

\begin{figure}[h]
\centering{
\includegraphics[scale=0.55,clip=]{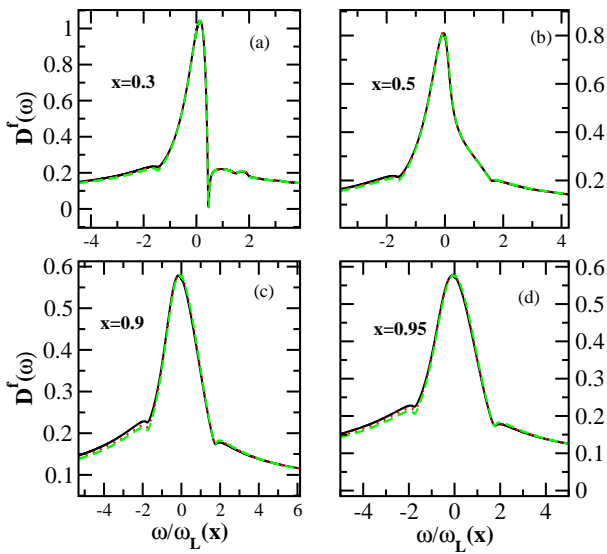}
}
\caption{(color online) Local f$-$ spectral function varying with scaled frequency $\om/\om_L(x)$ for substitution values $x=0.3$, $x=0.5$, $x=0.9$ and $x=0.95$. The model parameters are the same as in figure~\ref{fig:unv1}.
}
\label{fig:unv}
\end{figure}

The clean limit of the PAM has been studied extensively~\cite{grenz1,jarrell,roze,grew,newns,vidh03,vidh04,vidh05}. It has been found~\cite{vidh03,vidh04,vidh05} that the spectral functions, optical conductivity and resistivity in the strong coupling regime are universal functions of $(T/\om_L, \om/\om_L)$. The low energy scale, which is given by $\om_L \simeq ZV^2/t_*$ where $Z=(1-\partial\Sigma/\partial\omega|_{\om=0})^{-1}$ is an exponentially decreasing function~\cite{loga98,vidh04} of $U/V^2$ (for $\eta=1+2\eps_f/U=0;\ U/V^2\gg 1$). Substituting $f$-electrons with Kondo holes should give rise to significant changes in the local $f$-electron spectrum and the low energy Kondo scale. In a recent work~\cite{pram13}, we have shown that the CPA self-energy develops a finite linear in frequency imaginary part, thus a definition of quasiparticle-weight using the CPA self-energy is not possible. However, the local self-energy does have adiabatic continuity to the non-interacting limit, and hence may be used to define a low energy scale, which would naturally depend on the Kondo hole concentration, $x$. We define a low energy scale, $\om_L(x)$ as $Z(x)V^2/t*$, where $Z(x)$ is the quasiparticle weight of the local self-energy for a given $x$. For Kondo hole substituted systems in the strong coupling limit, the low energy scale $\om_L(x)$ is exponentially small, which is a prerequisite for the scaling consideration of spectral quantities. In figure ~\ref{fig:unv1}, we have shown $\om_L(x)$, which is indeed exponentially decreasing with increasing Coulomb interaction $U$ for different concentrations.
Further, in figure~\ref{fig:unv}, we show the universal behaviour of the local f$-$ spectral function $D^f(\om)$ for different substitution values.
In the panel (a) of figure~\ref{fig:unv}, we show the local f$-$ spectral functions {\it vs} scaled frequency $\om/\om_L(x)$ for different Coulomb interactions $U=6.2$ (solid line), $7.2$ (dotted line), $8.2$ (dashed line) for substitution value $x=0.3$. The f$-$ spectral functions for different $U$ collapse onto a universal form. Similarly in panels (b), (c) and (d) the scaling of f$-$ spectral functions have been shown for the $x=0.5$, $x=0.9$ and $x=0.95$ respectively. Such universal behaviour of the local f$-$ spectral functions for a wide range of substitution values concludes the presence of a low energy scale for Kondo hole substituted heavy fermions. 
Nevertheless, it is important to note that such universal scaling
is not obtained for the disorder averaged, i.e the CPA Green's functions. This naturally implies that transport or other quantities,
that depend on the CPA Green's functions will
not exhibit a scaling collapse as a function of varying interaction
strength. 

\begin{figure}[h]
\centering{
\includegraphics[scale=0.65,clip=]{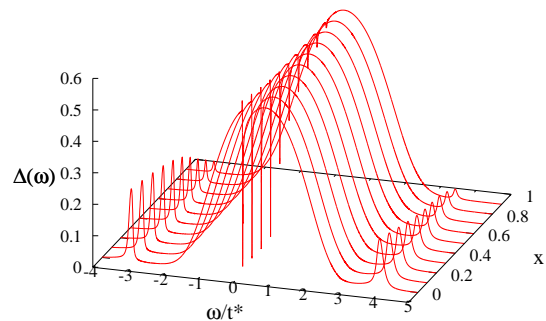}
}
\caption{(color online)Hybridization as a function of absolute frequency $\om/t_*$ for various substitution ($x$) 
values. The parameters are $U=5.23; V^2=0.4; n_f=0.98; n_c=0.53$.
}
\label{fig:dos_5}
\end{figure}

 The hybridization function, $\Delta(\om)=-{\rm Im}\left[S(\om)\right]$, (where the $S(\om)$ is the Feenberg self-energy) depends, naturally, on Kondo hole concentration. This $\Delta(\om)$ may be found through the imaginary part of the inverse of the host Green's function, which is determined self-consistently within DMFT~\cite{geor}. We show the $\Delta(\om)$ in figure~\ref{fig:dos_5}. It is seen that for small values of concentrations, the hybridization function has a Gaussian envelope with spectral weight carved around $\epsilon_{f*}=Z(\eps_f+\Sigma(0))$. With increasing concentration, the hybridization gap fills up and in the single impurity limit, ($x \simeq 1$) we see a featureless Gaussian. This is expected, because in the dilute limit, the impurities should have a negligible effect on the host, hence the hybridization assumes a simple form that is proportional to the non-interacting density of states, which has been chosen to be a Gaussian in our work.

\begin{figure}[h]
\centering{
\includegraphics[scale=0.40,clip=]{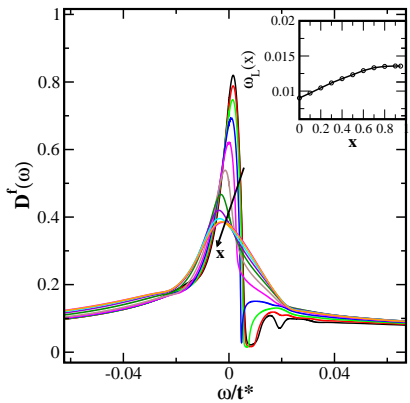}
}
\caption{(color online) Expanded view of the low frequency region of the local $f$-dos. Inset: variation of low energy scale with Kondo hole concentration.
}
\label{fig:wl_5}
\end{figure}

   One important inference can be made here about the difference between the influence of Kondo hole disorder {\it vs.} temperature. The spectral weight transfer into the hybridization gap is seen to arise from high energy scales, even from the Hubbard bands (figure~\ref{fig:dos_5}). Thus, disorder is seen to affect all energy scales. Temperature, in contrast, affects the spectrum only on
energy scales that are comparable to the thermal energy scale~\cite{vidh05}. Hence the incoherence effects induced by disorder are quite distinct to those by temperature.
In the figure~\ref{fig:wl_5}, we show the low frequency region of the local $f$-dos, $D^f(\om)=-{\rm Im}G^f(\om)/\pi$, (equation~\ref{eq:locgf}) as a function of `bare' frequency, $\om/t_*$ for various values of the Kondo hole concentration, $x$. It is easy to see that a redistribution of spectral weight has occurred with the increase in $x$, and the hybridization gap flanking the Kondo resonance fills up giving rise to a broad  resonance in the single-impurity limit. The full-width at half-maximum of the resonance is expected to be proportional to the low energy scale. And given the broadening of the resonance, we must expect that the $\om_L$ should increase with increasing $x$. Indeed, as the inset shows, the $\om_L$ rises almost linearly, and saturates in the single-impurity limit. 

 Next, we will discuss the effect of disorder on  finite temperature static response functions i.e resistivity, thermopower, Hall coefficient and Hall angle. 

\subsection{DC Resistivity}
\label{subsec:res_5}
 
\begin{figure}[h]
\centering{
\includegraphics[scale=0.45,clip=]{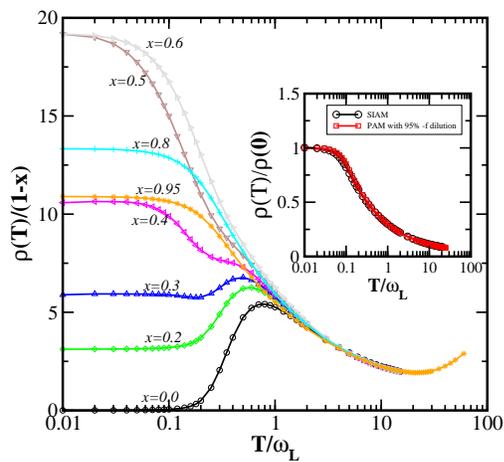}
}
\caption{(color online) Main panel: Resistivity per $f$- site  as a function of scaled temperature, $T/\om_L$ (model parameters are $U\simeq 5.11, V^2=0.6, n_f\simeq 0.98, n_c\simeq 0.59$). Inset: The strong coupling single impurity Anderson model (SIAM) resistivity compared with concentration value $x=0.95$. Model parameters for SIAM are $U=5.11; V^2=0.2; \eps_c=0.5$.
}
\label{fig:dcres_5}
\end{figure}
In the main panel of figure~\ref{fig:dcres_5}, the effects of Kondo hole substitution on DC resistivity $\rho(T)$ {\textit vs} scaled temperature $T/\om_L$, where $\om_L$ is the low energy scale for $x=0$, have been shown. For zero concentration, resistivity is zero at $T=0$ and follows $T^2$ behavior (Fermi-liquid) at low temperatures. As temperature is increased, a crossover from coherent to incoherent behavior in resistivity take place. At high temperatures ($T \gg \omega_L$), the resistivity shows the asymptotic single impurity Hamann form ($\rho(T)=\frac{3\pi^2}{16\ln^2{(T/\om_L)}}$) as discussed in detail in previous work ~\cite{vidh05}. The presence of a coherence peak signifies the crossover at low temperatures to coherent lattice behaviour. Coherence peak shifts to lower temperature value with increasing Kondo hole concentration. Since the resistivity decreases monotonically with increasing temperature for $x\gtrsim 0.4$, lattice coherent behaviour never sets in for the higher concentration values.  At $T/\om_L \gg 1$, the resistivity for all $x$ collapses onto a single universal form, which is simply the resistivity for a single-impurity Anderson model (see inset of figure~\ref{fig:dcres_5}~\cite{costi}). The residual resistivity (not shown) does not follow Nordheim's rule ($\rho(T=0)\propto x(1-x)$), which is consistent with previous work ~\cite{grenz} and experiments~\cite{sato}.

\begin{figure}[t]
\centering{
\includegraphics[scale=0.6,clip=]{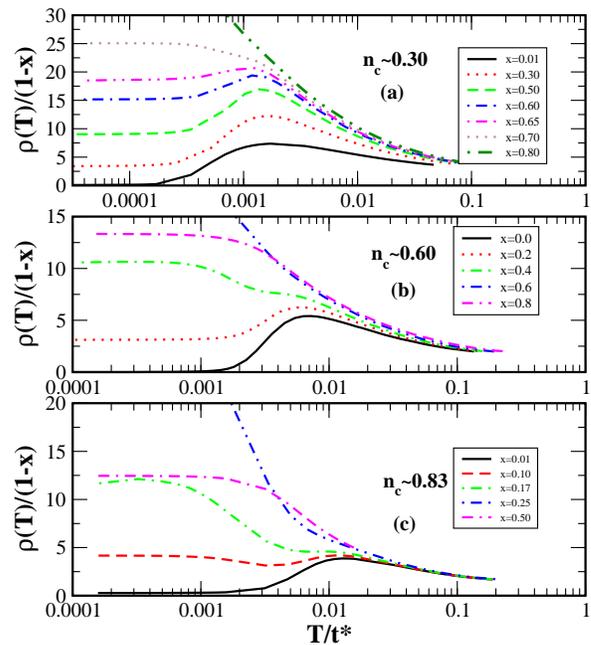}
}
\caption{Resistivity as function of temperature for three conduction electron occupancies and various concentration values. The coherence peak is seen to disappear beyond $x\gtrsim 1-n_c$, which are roughly $0.7, 0.4$ and $0.2$  for $n_c\sim 0.30$ (top panel), $n_c\sim 0.60$ (middle panel) and $n_c\sim 0.83$ (bottom panel) respectively.The model parameters are $V^2=0.6$; $U\sim 5.20$.}
\label{fig:crossover}
\end{figure}

      In a few recent works, the authors~\cite{kaul,fulde} used CPA combined with slave-Boson mean field , to show that a `critical' concentration of $x\sim (1-n_c)$ is required to induce a crossover from lattice coherent behaviour to single impurity incoherent behaviour. This implies that the crossover to incoherence is dependent on the conduction electron concentration. For symmetric Kondo insulators, since the $n_c=1$, this crossover would occur for an infinitesimal concentration of Kondo holes, while in the exhaustion regime~\cite{nozier}, the crossover would require a high substitution of the non-magnetic homologue. We investigate this conduction electron dependence in the crossover through a study of the coherence peak in the resistivity shown in figure ~\ref{fig:crossover}. 

               In top panel (a) of figure~\ref{fig:crossover}, we show the DC resistivity for $n_c\sim 0.3 $. The coherence peak is present upto $x\sim 0.65$ and resistivity follows single impurity behaviour beyond. Similarly in middle panel(b) and bottom panel (c) of figure~\ref{fig:crossover}, crossover from lattice coherent to single impurity incoherent behaviour takes place at $x\sim0.4$ and $x\sim0.17$ for $n_c\sim0.6$ and $n_c\sim0.83$ respectively. Thus, our results are consistent with the finding in Ref $22$, mentioned in the above paragraph.   

\subsection{Thermopower}
\label{subsec:therm_5}
\begin{figure}[h]
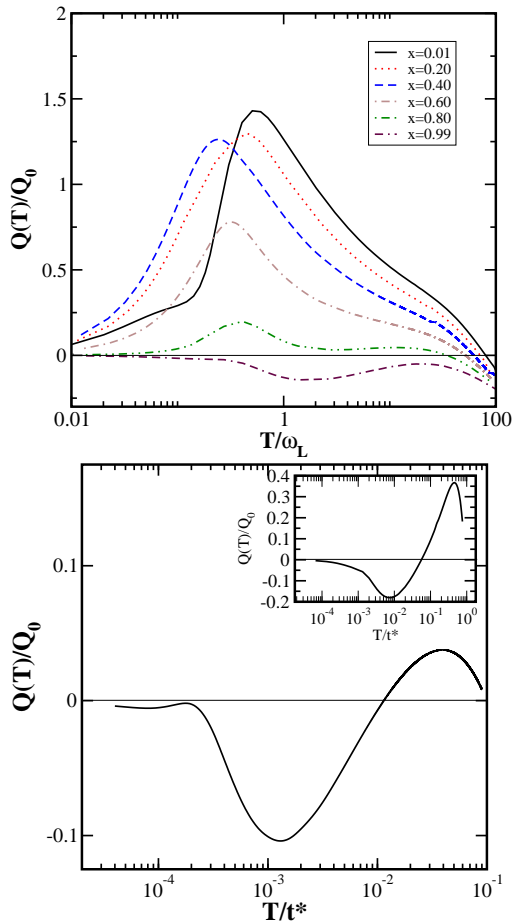

\centering{
\includegraphics[scale=0.45,clip=]{fig7a.eps}
\includegraphics[scale=0.46,clip=]{fig7b.eps}
}
\caption{(color online) Upper panel: Thermopower {\it vs.} $T/\om_L$. (model parameters are $U\sim 5.23, V^2=0.4, n_f\sim 0.98; n_c\sim 0.53$). Lower panel: Thermopower {\it vs.} $T/t^*$ in dilute limit i.e $x=0.99$ with $U\sim 7.17, V^2=0.4, n_f\sim 0.97, n_c\sim 0.52$. Inset: Thermopower of SIAM  for $U\sim 5.80 ;V^2=0.4$.
}
\label{fig:dctherm_5}
\end{figure}
The effect of Kondo hole substitution on thermopower for different temperatures has been shown in the upper panel of figure~\ref{fig:dctherm_5}. Like resistivity, temperature has been scaled by low energy scale of $x=0$. In the clean case, the thermopower rises from zero, reaches a maximum at a universal temperature, $T\sim \om_L$, and subsequently decreases monotonically, with a change of sign at non-universal temperatures. This functional form is preserved for almost all $x$, with a distinct form arising only in the extreme dilution limit ($x\rightarrow 1$). However, the position of the `coherence peak' exhibits an interesting feature with varying $x$, that is related to the `critical' $x$ at which the crossover from coherent lattice to single impurity incoherent behaviour occurs in the resistivity. For $x\lesssim 0.5$, the position of the maximum in thermopower red shifts monotonically with increasing concentration of Kondo holes, and for higher $x$, begins to blue shift (upper panel of figure~\ref{fig:dctherm_5}). The magnitude of this peak however decreases monotonically with increase in $x$ and changes sign in the single-impurity limit. In the extreme dilution limit shown in the lower panel of figure~\ref{fig:dctherm_5}, the thermopower looks qualitatively similar to that of SIAM~\cite{costi} (inset of lower panel), i.e. one peak at low temperature and the other peak with opposite sign at large temperature.
\subsection{Hall coefficients and Hall Angle}
\label{subsec:hall}

\begin{figure}[h]
\centering{
\includegraphics[scale=0.50,clip=]{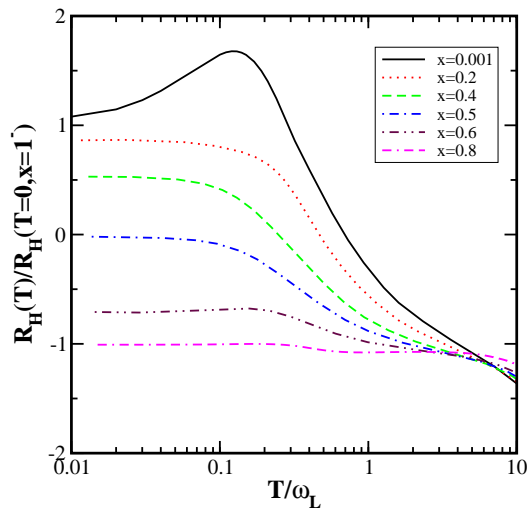}
}
\caption{(color online) Hall coefficient $R_H(T)$ {\it vs.} temperature $T$ scaled by low energy scale $\om_L(x=0)$. The model parameters are $U\sim 5.35 ;V^2=0.6$ and $\ep_c=0.5$ for which the occupancies are $n_f~1.0$ and $n_c~0.55$.}
\label{fig:hall}
\end{figure}

The Hall coefficient, $R_H$, in conventional metals is temperature independent, and a simple measure of the carrier type and density. Heavy fermion metals, on the other hand, exhibit a highly temperature dependent and material specific Hall coefficient $R_H$~\cite{onuki2}. Various theoretical explanations for anomalous Hall effect have been discussed in detail in the recent review by S. Nair {\it et. al}~\cite{nair}. In the figure~\ref{fig:hall}, we show the Hall coefficient (scaled by the $R_H$ at $T=0$ of the single impurity) {\it vs.} scaled temperature ($T/\om_L$), where $\om_L$ is low energy scale for zero Kondo hole concentration, for various values of the Kondo hole concentration, $x$. In the clean Kondo lattice limit ($x\rightarrow 0$), the Hall coefficient has a finite positive value, which increases with increase in temperature, peaks around $T\sim0.5 \om_L$ and then decreases monotonically with a  change of sign at higher non-universal  temperatures. At zero temperature, the $R_H$ decreases in magnitude and eventually as $x\rightarrow 1^-$, changes sign with increasing $x$. In parallel to the behaviour in resistivity, a collapse of $R_H$ {\it vs.} $T$ is found at higher temperatures ($T\gg\om_L$) for all $x$ reflecting a crossover from lattice coherent behaviour to single-impurity behaviour, as a function of $T$ and $x$.
Since we have computed the resistivity and the Hall coefficient, it is straightforward to explore the Hall angle, which is defined as $\theta_H =\cot ^{-1} \left(\rho(T)/R_H(T)\right)$, as a function of $x$ and $T$. Since the $R_H$ changes sign with increasing $T$ for $x\lesssim 0.4$, we expect, in this range of $x$, the Hall angle to show sign change with increase in temperature.

\begin{figure}[h]
\centering{
\includegraphics[scale=0.50,clip=]{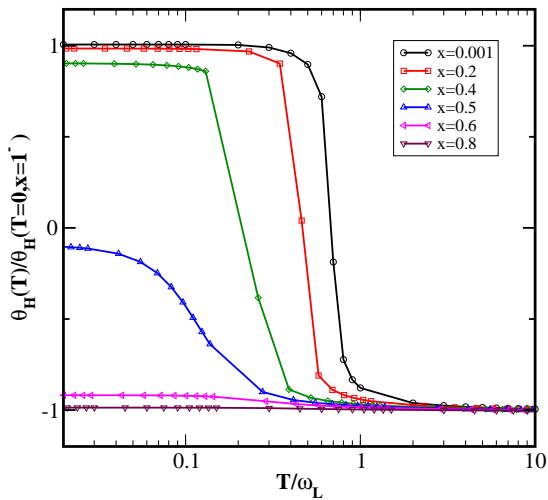}
}
\caption{(color online) Hall angle {\it vs.} temperature $T$ scaled by low energy scales $\om_L$ at $x=0$. (model parameters are same as for figure ~\ref{fig:hall}).}
\label{fig:hallangle}
\end{figure}

    In figure~\ref{fig:hallangle}, the variation of Hall angle with temperature has been shown for different concentrations of Kondo holes. In the concentrated limit (below $x \lesssim 0.4$), Hall angle has finite positive value at low temperature and changes sign sharply at large non-universal temperatures. The sign change occurs only for $x\lesssim 1-n_c$ and beyond that, the sign of Hall angle does not change. An important fact to be noticed here is that the sign change in Hall angle occurs almost like a first order transition, which is in complete contrast to the smooth crossover seen in resistivity and Hall coefficient, which are numerator and denominator respectively of the Hall angle ($\theta_H =\cot ^{-1} \left(\rho(T)/R_H(T)\right)$).
        
               In the next subsection, we discuss the effects of Kondo hole substitution on dynamical response functions. We consider optics first.
\subsection{Optical conductivity and optical scattering rate}
\label{subsec:opt_5}

\begin{figure}[h]
\centering{
\includegraphics[scale=0.50,clip=]{fig10a.eps}
\includegraphics[scale=0.50,clip=]{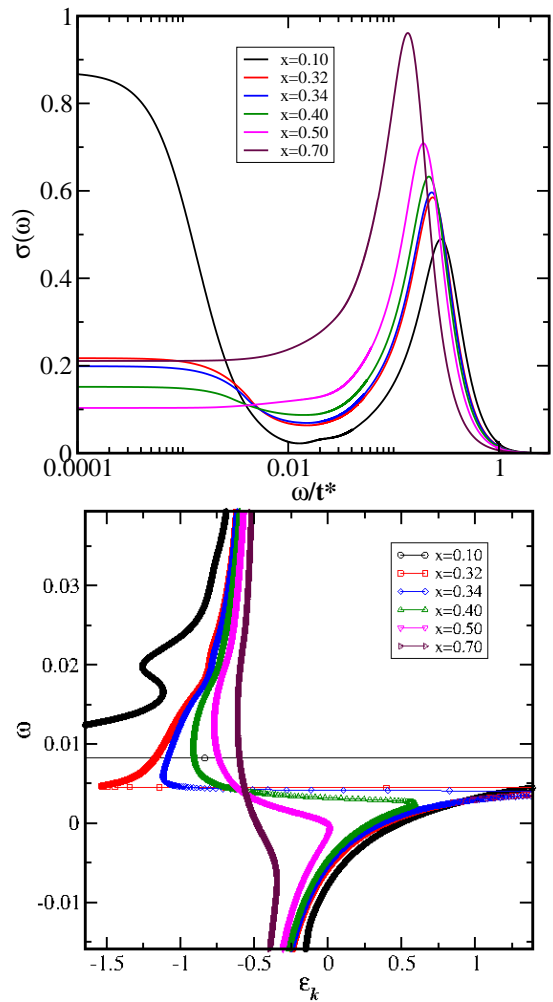}
}
\caption{(color online) Top panel: Zero temperature optical conductivity as a function of $\om/t*$ for various Kondo hole concentrations.
Bottom panel: Band dispersion for various $x$ values. The model parameters are $U=5.11; V^2=0.6 ;n_f\simeq 0.98$ and $n_c\simeq 0.59$. }
\label{fig:opt_5}
\end{figure}
\begin{figure}[h]
\centering{
\includegraphics[scale=0.32,clip=]{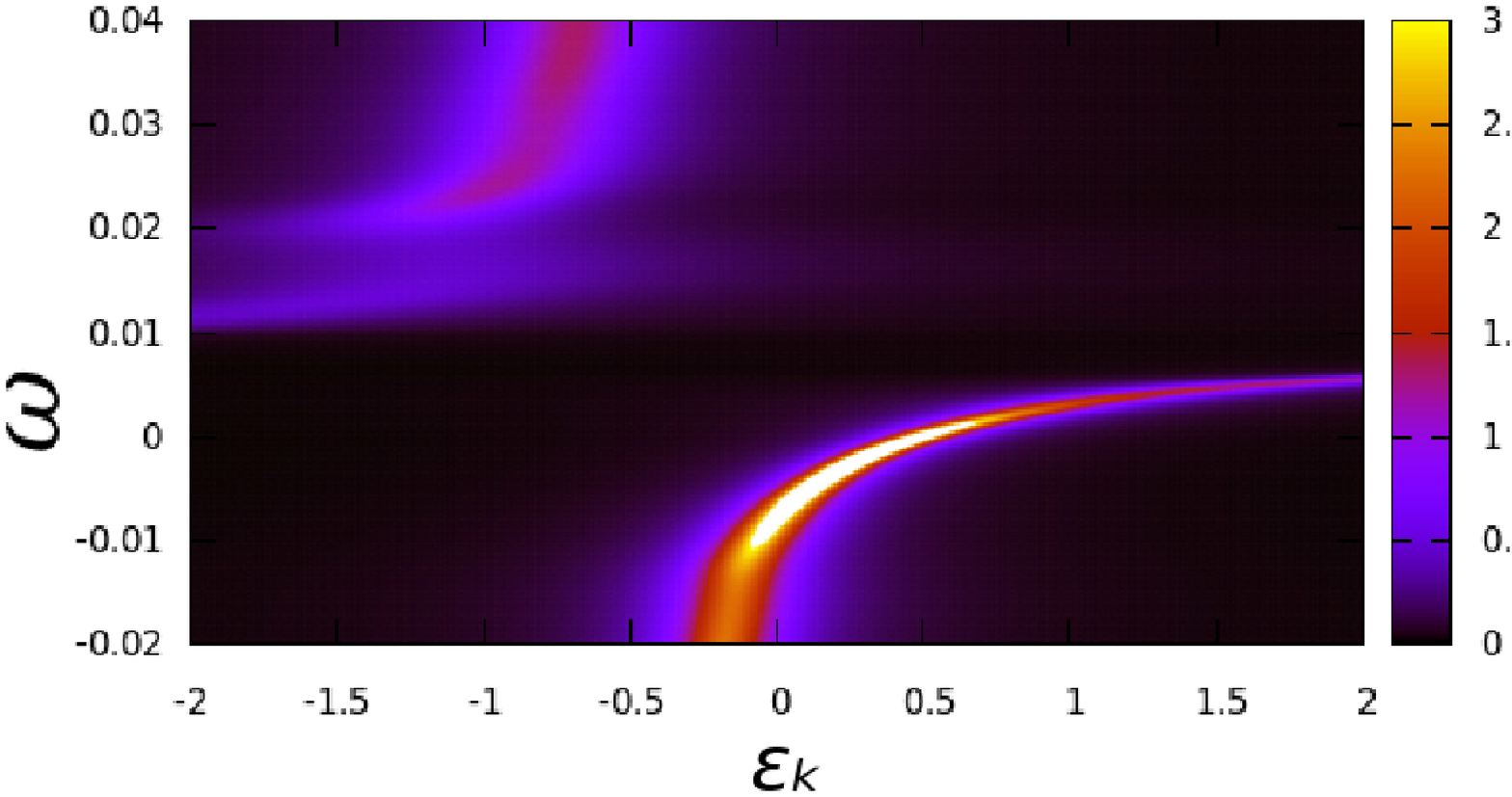}
\includegraphics[scale=0.32,clip=]{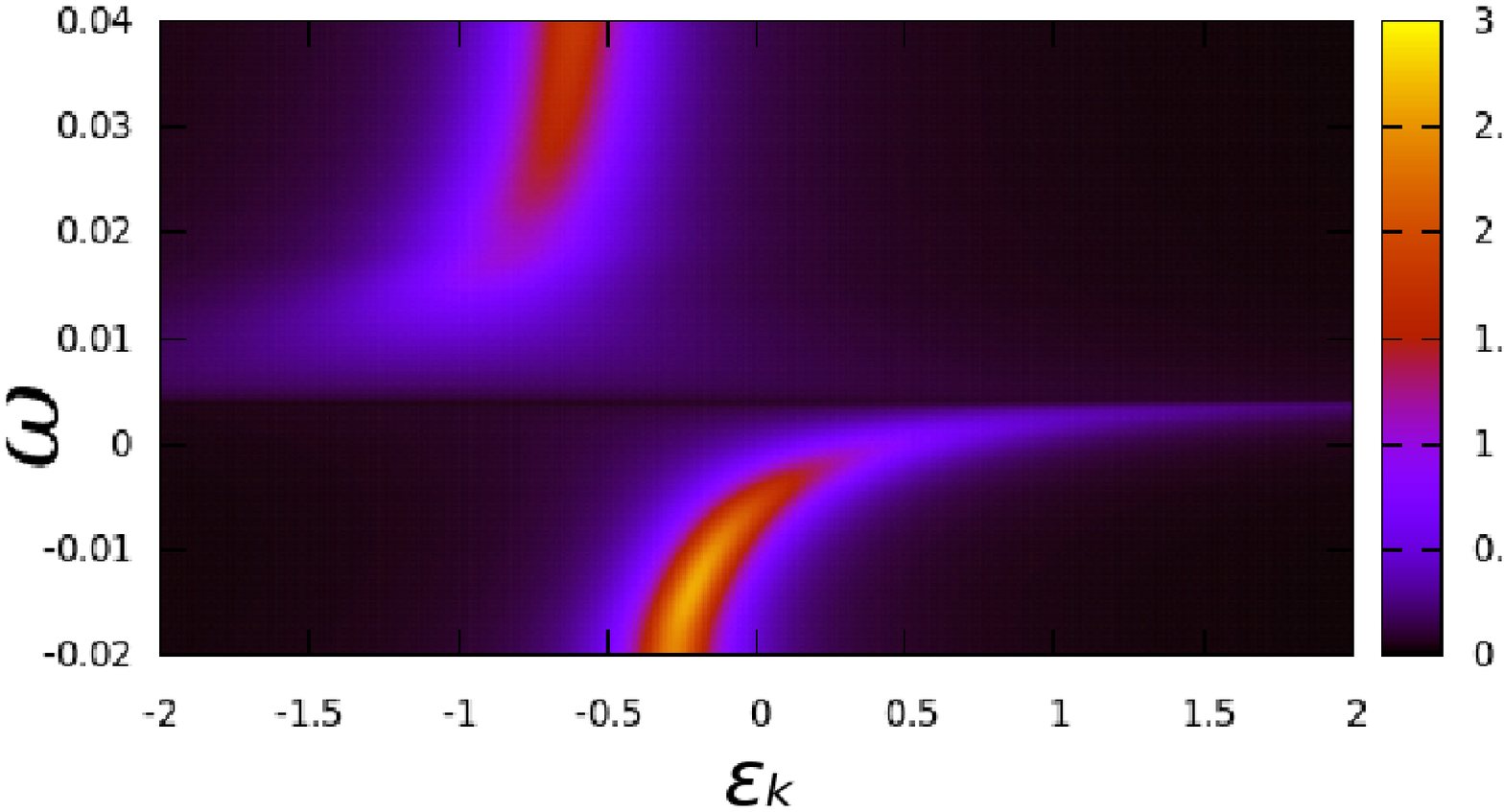}
}
\centering{
\includegraphics[scale=0.32,clip=]{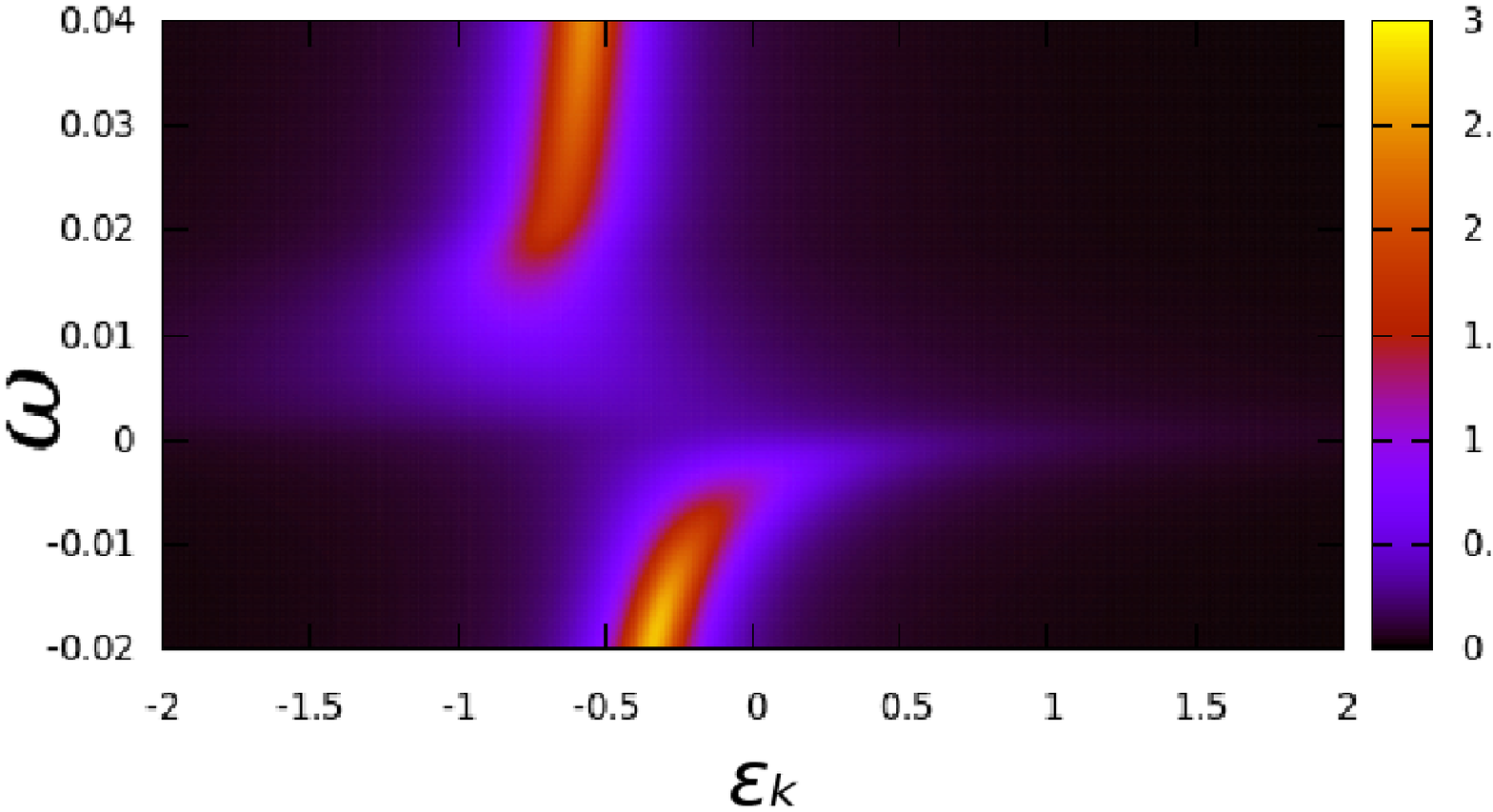}
\includegraphics[scale=0.32,clip=]{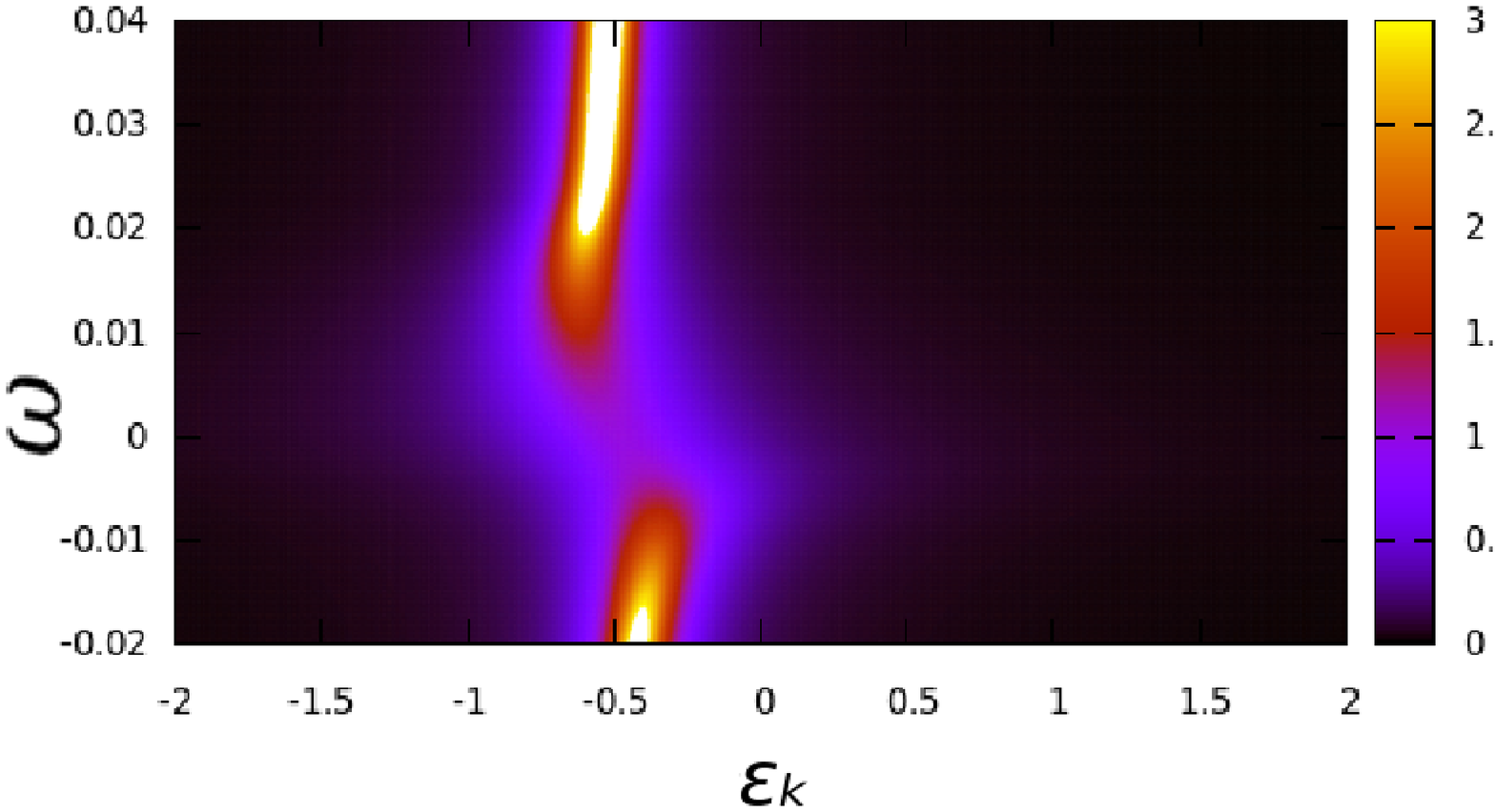}
}
\caption{ False colour contour plot of the single-particle dispersion,
 $D^{\CPA}(\ep_\k,\om)=-{\rm Im}G^{\CPA}_c(\ep_\k,\om)/\pi$ for concentration values $x=0.1$, $x=0.34$, 
$x=0.5$  and $x=0.7$ (from top to bottom). The model parameters are same as figure ~\ref{fig:opt_5}. 
}
\label{fig:optics1}
\end{figure}

In the left panel of figure~\ref{fig:opt_5}, we show the $T=0$ optical conductivity computed using equation~\ref{eq5_18} for different values of $x$. With increasing $x$, the Drude peak at $\om=0$ melts rapidly and the low frequency region appears flat and featureless. The DC value of the optical conductivity represents static effects of impurity scattering. The mid-infrared peak moves to lower frequencies with increase in Kondo hole concentration. This is counter-intuitive if we invoke the renormalized non-interacting picture, which says that the MIR peak is positioned at $\sim\sqrt{ZV^2}$. The scale increases with $x$, so if the MIR were to be proportional to $\sqrt{\om_L}$, then the MIR would experience a blue shift. So how does one explain the red shift? The answer is provided by the dispersion, $\om(\epsilon_\k)$ found by locus of zeroes of the ${\rm Re} \left[G^{\CPA}_c(\ep_\k,\om)^{-1}\right]$. This is shown in the right panel of Fig.~\ref{fig:opt_5}.  It is seen that for low concentration, there is a clean minimum direct gap, that is indeed proportional to $\sqrt{\om_L}$. With increasing Kondo hole concentration, the direct gap fills up with mid-gap states, which causes the gap to direct excitation to decrease. Eventually, for $x\gtrsim 0.7$, there is almost no gap. Thus, the theory predicts that with increasing substitutional disorder, the MIR absorption peak should experience a strong red shift. The imaginary part of the self-energy represents the damping of the quasiparticles, and the band structure shown in figure~\ref{fig:opt_5} does not fully capture this aspect, since only the real part of the denominator of the CPA Green's function is used. To remedy this, we also show the full band structure by plotting $\epsilon_\k$ and frequency dependent $-\Im G^{\CPA}(\om,\ep_\k)/\pi$ as a two dimensional contour plot (with false colours) in figure~\ref{fig:optics1} for four $x$ values. We observe that, at $x=0.1$, there is almost no spectral weight in the region between the two bands, implying that the MIR peak would be a prominent high energy feature. With increasing $x$, the two bands come closer and appreciable spectral weight appears in the form of mid-gap states arising due to Kondo holes. This indeed implies that the MIR peak will redshift and simultaneously, the absorption will be finite all the way from the peak down to $\om=0$. Thus, the incoherent scattering by random Kondo hole substitution is responsible for the red-shift of the MIR peak and the concomitant destruction of the Drude peak.  

\begin{figure}[h]
\centering{
\includegraphics[scale=0.45,clip=]{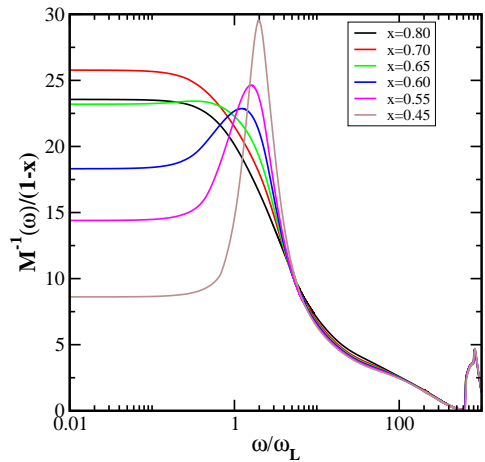}
}
\caption{(color online) Zero temperature optical scattering rate as a function of $\frac{\om}{\om_L}$ ($\om_L$ is low energy scale at $x=0$) 
 for various disorder strengths. The model parameters are $U=5.32; V^2=0.6 ;n_f\simeq 0.97$ and $n_c\simeq 0.43$.
}
\label{fig:optscat_5}
\end{figure}

The optical scattering rate, $M^{-1}(\om)$, defined in section~\ref{sec:model_5} as $M^{-1}={\rm Re} (1/\bar{\sigma}(\om))$, is shown in figure~\ref{fig:optscat_5}. In the concentrated regime ($x\rightarrow 0$), a characteristic peak is visible in the optical scattering rate at low frequencies. This is also observed in experiments~\cite{basov2,kimura2,okumura} on heavy fermion systems. 
This peak is narrow and centred at $\om_L$ for small $x$. As $x$ increases, the peak broadens, experiences a red shift, and ultimately vanishes in the dilute limit. It is precisely around $x\sim 0.6$, that this peak structure vanishes, which is attributed to crossover from heavy fermion to single impurity regime. The high frequency tail is seen to be universal for all $x$. We further investigate the effect of temperature on the optical scattering rate for finite value of Kondo hole concentration and temperature.
\begin{figure}[h]
\centering{
\includegraphics[scale=0.5,clip=]{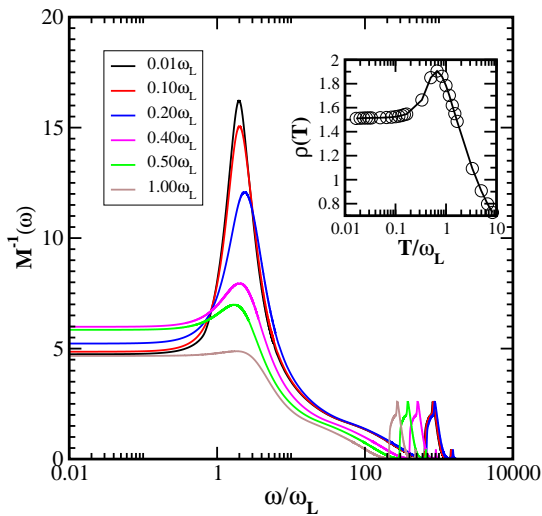}
}
\caption{(color online) Main panel: optical scattering rate for $x=0.45$ and various temperatures, shown as fractions of the low energy scale. In the inset, the resistivity {\it vs.} scaled temperature is shown, also for $x=0.45$. The other model parameters are $V^2=0.6; \eps_c=0.7 ; \eta\simeq 0$ and $U\simeq 5.32$,$n_f=0.97$ and $n_c=0.43$.}
\label{fig:optscattemp_5}
\end{figure}
In the main panel of figure~\ref{fig:optscattemp_5}, the optical scattering rate has been shown for $x=0.45$ versus scaled frequency $\om/\om_L$.
The peak in optical scattering rate corrodes slowly with increasing temperature, and finally vanishes for $T\gtrsim 0.5\om_L$ for the parameters mentioned in Fig.~\ref{fig:optscattemp_5}. In the inset of figure~\ref{fig:optscattemp_5}, the DC resistivity {\it vs.} temperature has been shown for the same parameter regime. It is seen that the coherence peak appears at the same value of temperature i.e $T\sim 0.5\om_L$, where peak in scattering rate vanishes (main panel) and for all higher temperatures, the resistivity follows single impurity behaviour. Thus, the behaviour of optical scattering rate is consistent with resistivity in terms of predicting the crossover from Kondo lattice (KL) to single impurity.
\section{Comparison to experiment}
\label{sec:exp_5}
\subsection{Resistivity}
\begin{figure}[h]
\centering{
\includegraphics[scale=0.47,clip=]{fig14a.eps}
\includegraphics[scale=0.35,clip=]{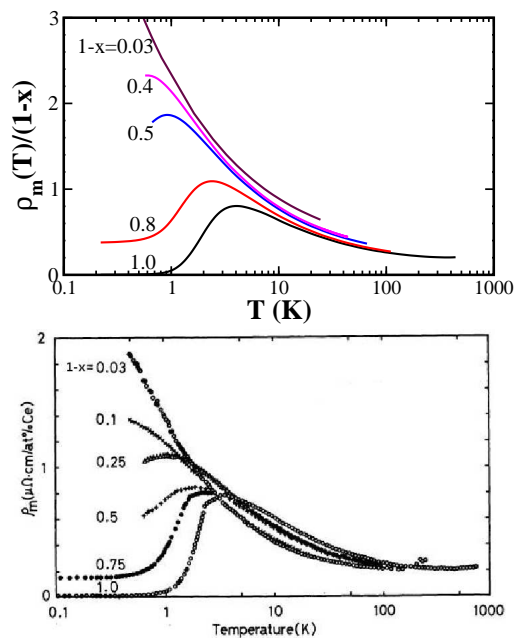}
}
\caption{(color online)Comparison of theory with experiment for Ce$_x$La$_{1-x}$. Left panel: theoretically computed resistivity versus $T$ for various $x$. In right panel, experimental data for Ce$_x$La$_{1-x}$B$_6$ by Sato \etal}
\label{fig:expres_5}
\end{figure}

In previous work ~\cite{vidh06}, DMFT+LMA  has been employed to compare theory with experiments for a few heavy fermion metals in the clean case. Theoretical comparisons with experiment for disordered case has several complications. Substitutional disorder may change lattice constants which effectively can change the hopping parameters, site energies  and hybridization amplitudes. A precise estimation of model parameters for different values of concentration is next to impossible and thus only qualitative comparison is possible. In figure~\ref{fig:expres_5}, we have compared concentration dependent resistivity of Ce$_x$La$_{1-x}$B$_6$ by N. Sato {\it et al} ~\cite{sato} with our theory. 
In the top panel, we present theoretical data where $n_f=0.98$, $n_c=0.53$ and $U/V^2\sim 6.0$. The bottom panel reproduces the (phonon subtracted) experimental data of Sato \etal~\cite{sato}. With the above choice of parameters and appropriate scaling (mentioned in previous work ~\cite{vidh06}), the theoretical data matches excellently with experimental data (right panel of figure~\ref{fig:expres_5}) for the clean case ($x=0$). If we compute resistivities for finite $x$ without changing the model parameters, we find that the residual resistivity peaks at a finite $x$, which contradicts the experimental observation that the residual resistivity increases monotonically with increasing $x$ and saturates in the dilute limit. Hence, in order to get correct trend in residual resistivity with increasing disorder, we introduce a minimal dependence of a single model parameter with $x$. Our choice is the linear dependence of $x$ for the conduction orbital site energy ($\ep_c(x)=\ep_c(0)+\alpha x$, with $\alpha=0.5$) which effectively determines the hybridization ($V^2/(\om-\ep_c(x)-S(\om)$) of f$-$ electrons with the conduction bath. The argument behind such a choice is that the larger atomic size of the doped lanthanum atom changes the effective hybridization. The argument is consistent with experimentally found increase in lattice constant upon Ce substitution with La~\cite{JLTP}. Further, the $x$-axis is scaled by ratio of coherence peak position in theory to the experiment for zero disorder. The agreement between theory and experiment is seen to be qualitatively good.

\subsection{Thermopower}
\begin{figure}
\centering{
\includegraphics[scale=0.40,clip=]{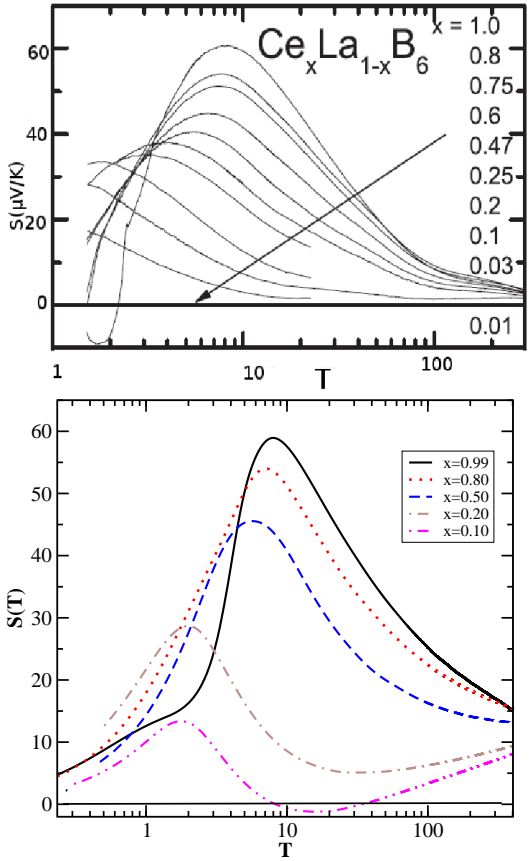}
\includegraphics[scale=0.47,clip=]{fig15b.eps}
}
\caption{(color online) Comparison of experiment with theory -- Top panel: Experimental data for Ce$_x$La$_{1-x}$B$_6$ by Kim {\it et al}~\cite{kuni}. Lower panel: Theoretically computed thermopower for various $x$.
}
\label{fig:exptherm_5}
\end{figure}

In the upper panel of figure~\ref{fig:exptherm_5}, thermopower measurement by Kim \etal~\cite{kuni} of Ce$_x$La$_{1-x}$B$_6$ for varying concentrations of Cerium is shown (note that the $x$ used in experiment is $1-x$ in our theory). The experimentally measured thermopower includes electronic ($f$) and lattice contributions. It is important to extract the electronic contribution in thermopower coefficient, since our calculation does not include phonons. For the case of DC resistivity, the Mattheissen's rule was employed to extract the electronic contribution. For thermopower, the Nordheim-Gorter rule $S \cdot \rho =S_{La}\cdot \rho_{La}+S_{Ce}\cdot\rho_{Ce}$ is commonly employed. The contribution from the first term is small and can be neglected (as argued in experimental work ~\cite{kuni}), thus $S_{Ce}=S$. It is observed that the peak position in thermopower shifts to lower temperatures with increasing $x$. In the lower panel of figure~\ref{fig:exptherm_5}, the theoretically computed thermopower is shown for the same parameter values as in figure~\ref{fig:expres_5}. The $x$-axis of theoretical data has  been scaled uniformaly for all $x$ by the ratio of the peak position in thermopower in theory to experiment for $x=0$. The theory does agree reasonably with experiments. Indeed it is gratifying to note that the theoretically computed DC resistivity and thermopower agree with experiments on La substituted CeB$_6$ for the same set of parameters.

\section{Conclusions}
In this paper, we have investigated Kondo hole type of substitution in heavy fermions using coherent potential approximation combined with dynamical mean field theory and local moment approach. The physics issue in focus is the crossover from heavy fermions to single impurity behaviour in physical properties like resistivity and thermopower. The approach used here does capture the crossover from Kondo lattice to single impurity behaviour as reflected in spectral functions, optics, resistivity, thermopower, Hall coefficient and optical scattering rate. The coherence peak in resistivity which is inherent to heavy fermion systems vanishes beyond a certain value of Kondo hole concentration. This value of concentration is dependent on conduction electron ($n_c$) filling. In the dilute limit, there is a sign change in thermopower. The zero temperature Hall coefficient and Hall angle also change sign at $x_c$. In the optical conductivity, Drude peak vanishes beyond the $x_c$. The peak structure in optical scattering rate and coherence peak in resistivity has one to one correspondence and are the measure of the coherence in the system. Comparison of our theoretical results with experimental data for resistivity and thermopower yields qualitatively good agreement. A concentration dependent conduction orbital energy correctly captures the experimental trend in  resistivity and thermopower. Coherent potential approximation does not capture inter-site coherence  and coherent back scattering effects. Recently developed approaches such as the typical medium-dynamical cluster approximation should be able to capture such effects and will be the subject of future investigation.

\end{document}